\documentclass[a4paper,12pt]{article}

\usepackage[T2A]{fontenc}
\usepackage{amsmath,amssymb}
\usepackage{cite}
\usepackage[linktocpage=true,plainpages=false,pdfpagelabels=false]{hyperref}

\newcommand{\ls}{\left(}
\newcommand{\rs}{\right)}
\newcommand{\la}{\lambda}
\newcommand{\dz}{\zeta}
\newcommand{\be}{\beta}
\newcommand{\three}[1]{\overset{{\scriptscriptstyle 3}}{#1}{}}

\newcommand{\disn}[2]{$$\displaylines{\refstepcounter{equation}%
            \label{#1}\hskip 1em minus 1em #2\hfilneg}$$}
\newcommand{\nom}{\hfil\hskip 1em minus 1em (\theequation)}

\newcommand{\ns}{\hfill\cr\hfill}

\begin{document}

\title{Canonical formulation of embedding gravity\\ in a form of General Relativity with dark matter}
\author{S.~A.~Paston\thanks{E-mail: pastonsergey@gmail.com},
T.~I.~Zaitseva\thanks{E-mail: taisiiazaitseva@gmail.com}\\
{\it Saint Petersburg State University, Saint Petersburg, Russia}
}
\date{\vskip 15mm}
\maketitle

\begin{abstract}
We study embedding gravity, a modified theory of gravity, in which our space-time is assumed to be a four-dimensional surface in flat ten-dimensional space. Based on a simple geometric idea, this theory can be reformulated as General Relativity with additional degrees of freedom and contribution to action, which can be interpreted as describing dark matter. We study the canonical formalism for such a formulation of embedding gravity. After solving simple constraints, the Hamiltonian is reduced to a linear combination of four first class constraints with Lagrange multipliers. There still remain six pairs of second class constraints. Possible ways of taking these constraints into account are discussed. We show that one way of solving the constraints leads to the canonical system going into the previously known canonical formulation of the complete embedding theory with an implicitly defined constraint.
\end{abstract}

\newpage

\section{Introduction}
The embedding approach to describe gravity was proposed by Regge and Teitelboim in \cite{regge}. The main object in this approach is some four-dimensional surface in a generally ten-dimensional flat ambient space with pseudo-Euclidean metric $\eta _{ab}$, described by the embedding function $y^a (x^\mu)$. Here $x^\mu$ are the coordinates on the surface; the function $y^a$ takes values in the ambient space; the Greek indices run through four values $0,1,2,3$, while the Latin ones run through ten values $0,1,\ldots ,9$. This surface is understood as our four-dimensional space-time. Its metric is induced by the embedding function, which means that the distance between infinitely close points of the surface is defined as the distance between these points in the ambient space:
\begin{equation}\label{induce}
    g_{\mu \nu} = (\partial _\mu y^a)(\partial _\nu y ^b) \eta _{ab}.
\end{equation}
The embedding gravity action (description of gravity as an embedding theory) is the usual Einstein-Hilbert action
\begin{equation}\label{EH_action}
     S = \frac{1}{2 \varkappa} \int d^4x \sqrt{-g} R + S_\text{m},
\end{equation}
in which the embedding function $y^a$ is chosen as the independent variable, while the metric is considered to be induced, that is, it is uniquely determined from the embedding function by (\ref{induce}). The term $S_\text{m}$ denotes the contribution of matter. In this work, we use the signature $-+++$.

Initially, the embedding theory was intended to solve some problems associated with the quantization of General Relativity. The fact is, the quantization procedure has proven itself well for field theories in a flat (pseudo-Euclidean) space. However, in the case of General Relativity, the space-time metric is a dynamic variable itself, and the known recipes stop working. A detailed review of the gravity quantization problems can be found in \cite{carlip}. In the embedding theory, however, there is a flat background metric, which can help in solving the problems of gravity quantization, so rewriting GR in this form may be useful.
After the pioneering work \cite{regge} embedding approach was analysed in \cite{deser}. In subsequent years the ideas of this approach were used repeatedly in the works of different authors to describe gravity, including in connection with its quantization, see, for example, \cite{tapia,maia89,estabrook1999,davkar,rojas09,statja25,faddeev,statja26}.

It is worth noting that embedding gravity is not equivalent to General Relativity due to the presence of the so-called extra solutions. Indeed, by varying the action (\ref{EH_action}) with respect to the independent variable $y^a$, one can obtain equations called the Regge-Teitelboim equations
\begin{equation}\label{sp1}
   D_\mu \Big( \big( G^{\mu \nu} - \varkappa\, T^{\mu \nu} \big) \partial _\nu y^a \Big) =0,
\end{equation}
where $G^{\mu \nu}$ is the Einstein tensor, $T^{\mu \nu}$ is the matter energy-momentum tensor, $D_\mu$ is the covariant derivative.
These equations are more general than the Einstein equations, that is, in addition to the solutions of the Einstein equations, they also contain the so-called <<extra>> solutions.
One can get rid of extra solutions by artificially imposing the so-called Einstein constraints \cite{regge}, which are four out of ten Einstein equations:
\begin{equation}\label{sp2}
G^{\mu 0} - \varkappa\, T^{\mu 0} =0.
\end{equation}
One can show \cite{statja18} that, being imposed at the initial moment, they are further satisfied due to the equations of motion \eqref{sp1}.

Later, extra solutions began to be considered not as a disadvantage of embedding gravity, but, on the contrary, as a virtue.
The reason for this was some interesting property of the embedding theory.
It turns out that the Regge-Teitelboim equations can be rewritten in the form corresponding to the Einstein equations with the contribution of some additional (fictitious) matter with the energy-momentum tensor $\tau ^{\mu \nu}$:
\begin{eqnarray}
 &&   G^{\mu \nu} = \varkappa \Big( T^{\mu \nu} +\tau ^{\mu \nu} \Big), \label{einst-extra} \\
 &&   D_\mu (\tau ^{\mu \nu} \partial _\nu y^a) = 0, \label{embedding_matter_motion}
\end{eqnarray}
this was first noticed in \cite{pavsic85}.
In the emerging fictitious \textit{embedding matter}, one can try to see \textit{dark matter}, the riddle of which is one of the unsolved problems of modern theoretical physics.
As a result, the dark matter phenomenon is explained as a purely gravitational effect, while the equation (\ref{embedding_matter_motion}) plays the role of the embedding matter equation of motion.
To analyse the properties of embedding matter, it may be convenient to equivalently reformulate embedding gravity as General Relativity with some additional contribution to the action \cite{statja48}.
Various possibilities for writing down this contribution are analysed in \cite{statja51}.
In particular, it is possible to obtain the nonrelativistic limit for the equations of embedding matter motion \cite{statja68}.

When trying to quantize a theory, an important step is to construct its canonical description.
Since when substituting \eqref{induce} into the action \eqref{EH_action} it contains the second time derivatives of the independent variable $y^a$, special methods \cite{tapia,rojas06,rojas09} can be used for canonical description construction.
However, one can see that after discarding the surface term in action, it can be written in a form containing only the first derivatives. Therefore, one can use the usual canonical approach.
For an embedding theory with an additional imposition of Einstein constraints \eqref{sp2}, the construction of a canonical formalism in this way was started in the pioneering work \cite{regge} and completed in the works \cite{statja18,statja24}.
The Hamiltonian of the theory turns out to be a linear combination of first class constraints.
When the ideal-forming part of the emerging constraints is satisfied, the complete constraint algebra goes \cite{statja24} into the well-known ADM constraint algebra \cite{adm}.

However, without imposing Einstein constraints, i.e. for full embedding gravity, the construction of the canonical formalism becomes much more complicated.
The reason for this is the fact that one of the constraints cannot be written explicitly, since this requires solving a multidimensional cubic equation, see details in \cite{statja44}.
Writing this "lapse" constraint was also discussed in \cite{frtap}.
Nevertheless, by carrying out the analysis in the presence of an implicitly defined constraint, one can show that in this case the Hamiltonian of the theory also reduces to a linear combination of the first class constraints, and also find the form of the algebra formed by these four constraints \cite{statja44}.
An interesting question is a connection between the constructed canonical description of full embedding gravity and the previously found canonical formalism with an additional imposition of Einstein constraints, but we are unable to trace this connection due to the complexity of the resulting expressions.

In this paper, we explore the canonical description of embedding gravity in the above formulation in the form of General Relativity with some additional contribution to the action, corresponding to embedding matter.
Such a canonical description is useful for the embedding matter properties better understanding.
In particular, when analysing the nonrelativistic limit of embedding matter motion equations, it is useful to classify the equations of motion according to the number of differentiations with respect to time, i.e. as constraints or dynamical equations, while in the canonical description such a classification occurs automatically.

In section \ref{eg_as_gr_add_matter}, we reformulate the embedding theory as General Relativity with embedding matter at the level of action.
By analogy with the Arnowitt-Deser-Mizner \cite{adm} variables, new variables for embedding matter description are introduced that are more convenient for the Hamiltonian formalism construction.
In section \ref{can_vars}, we move to the Hamiltonian description of the system and calculate all primary and secondary constraints.
The section \ref{classify} discusses whether the resulting constraints belong to the first or second class, and also solves some of the constraints that turn out to be quite simple.
Possible ways of further work with the remaining second class constraints are also discussed.
Their solution on the way of elimination of metric variables is considered in section \ref{compare}.
It is shown that, as a result, the canonical description of the complete embedding theory obtained earlier in \cite{statja44} is reproduced.

\section{Embedding gravity as GR with additional matter}\label{eg_as_gr_add_matter}
As mentioned in the Introduction, the equivalence between embedding theory and GR with fictitious embedding matter at the level of equations of motion allows one to achieve the same equivalence at the level of action.
That is, one can rewrite the action of embedding theory in the form
\begin{equation}\label{sp3}
    S = S^\text{EH} + S^\text{add}+ S_\text{m},
\end{equation}
where $S^\text{EH}$ is the Einstein-Hilbert action depending on the metric $g_{\mu \nu}$, which is one of the independent variables, and $S^\text{add}$ is some additional contribution containing, in addition to the metric, other independent variables that describe the embedding matter.
The simplest choice for this contribution is \cite{statja48}:
\begin{equation}\label{action}
    S^\text{add} =  \frac{1}{2} \int d^4 x \sqrt{-g} \Big(  g_{\mu \nu}-(\partial _\mu y^a)( \partial _\nu y_a) \Big) \tau ^{\mu \nu}.
\end{equation}
In this case, the embedding matter is described by the independent variables $y^a$ and $\tau ^{\mu \nu}$. The $\tau ^{\mu \nu}$ tensor is assumed to be symmetric.

One can easily check that varying the action \eqref{sp3} with respect to $\tau ^{\mu \nu}$ gives the condition of the induced metric \eqref{induce} while varying with respect to the metric $g_{\mu \nu}$ gives the Einstein equation \eqref{einst-extra}, and the independent variable $\tau ^{\mu \nu}$ gets the meaning of the embedding matter energy-momentum tensor.
To obtain a complete set of equations of motion, one still needs to vary with respect to the remaining independent variable $y^a$. This leads to the equation \eqref{embedding_matter_motion}.
As a result, one can see that the equations of motion corresponding to the action \eqref{sp3} are exactly equivalent to the Regge-Teitelboim equations.
In what follows, we will omit the contribution of ordinary matter $S_\text{m} $; if necessary, it can be restored easily.

It is well known using the Arnowitt-Deser-Mizner (ADM) variables \cite{adm}
\begin{equation}
\beta _{ik}=g_{ik}, \qquad
N_k=g_{0k}, \qquad
N=\frac{1}{\sqrt{-g^{00}}}
\end{equation}
simplifies the canonical formalism for GR.
The determinant of the three-dimensional metric $\beta _{ik}$ will be denoted as $\beta$.
Its connection with the determinant $g$ of the four-dimensional metric $g_{\mu \nu}$ is known: $\sqrt{-g} = N \sqrt {\beta}$.
Let us introduce the following notation: we will write the number $3$ over the quantities describing the three-dimensional internal geometry of the surfaces $x^0=const$.
For example, $\three{R}$ stands for the scalar curvature of three-dimensional space with the metric $\beta _{ik}$.
Such quantities depend only on $\beta _{ik}$, and, as a consequence, do not depend on $N$, $N_k$ or on derivatives $\partial _0 \beta _{ik}$.
We will also use the notation
\begin{equation}
e^a _\mu=    \partial _\mu y^a,
\end{equation}
calling this quantity the non-square vielbein.
From a non-square vielbein, one can compose longitudinal and transverse projectors for the space tangent to the surface $y^a(x^\mu)$ at a given point:
\begin{equation}
    {\Pi _\parallel} ^a _b = e^a _\mu e_{b} ^\mu , \qquad {\Pi _\perp} ^a _b = \delta ^a _b - e^a _\mu e_{b} ^\mu .
\end{equation}
However, three-dimensional projectors will be more useful for us:
\begin{equation}
    {\three{\Pi} _\parallel}{} ^a _b = e^a _i e_{bj} \widetilde{\beta} ^{ij} , \qquad {\three{\Pi} _\perp}{} ^a _b = \delta ^a _b - e^a _i e_{bj} \widetilde{\beta} ^{ij} ,
\end{equation}
where $\widetilde{\beta} _{ij} \equiv e ^a _i e_{a,j} $ is the three-dimensional induced metric, and $\widetilde{\beta} ^{ij}$ is its inverse. Here and below $i,j,\ldots=1,2,3$.
Note that in the described approach, the embedding function and the metric are independent variables, so $\widetilde{\beta} ^{ij}$ generally does not coincide with $\beta ^{ij}$. They start to coincide only on-shell.

By discarding the surface terms, the Einstein-Hilbert action $S^\text{EH}$ can be rewritten in a form called the ADM action  \cite{adm}:
\begin{equation}
S^\text{ADM}=
\int d^4x \Big( 2N K_{ik} L^{ik,lm} K_{lm} + \frac{1}{2\varkappa} N \sqrt{\beta} \three{R} \Big) .
\end{equation}
Here the quantity
\begin{equation}
    K_{ik} = \frac{1}{2N} \Big( \three{D} _i N_k + \three{D} _k N_i - \partial _0 \beta _{ik} \Big)
\end{equation}
is called the second quadratic form, and
\begin{equation}
    L^{ik,lm} = \frac{\sqrt{\beta}}{8\varkappa} \Big( \beta ^{il} \beta ^{km} + \beta ^{im} \beta ^{kl} - 2 \beta ^{ik} \beta ^{lm} \Big)
\end{equation}
is called the Wheeler-de-Witt metric.
The metric inverse to it has the form
\begin{equation}
       \overline{L} _{ik,lm} = \frac{2 \varkappa}{\sqrt{\beta}} \Big( \beta _{il} \beta _{km} + \beta _{im} \beta _{kl} - \beta _{ik} \beta _{lm} \Big),
\end{equation}
so that
\begin{equation}
    \overline{L} _{ik,lm} L^{lm,pr} = \frac{1}{2} \Big( \delta ^p _i \delta ^r _k + \delta ^p _k \delta ^r _i \Big).
\end{equation}

Just as ADM variables simplify calculations when constructing the canonical formalism for General Relativity, in this theory it is convenient to pass from the variables $\tau ^{\mu \nu}$ to the new variables $\phi$, $\phi ^k$, and $ \phi ^{ij}$:
\begin{eqnarray}
&& \phi= - \frac{1}{2} N^2 \sqrt{\beta} \tau ^{00}   ,\\
&& \phi ^k =  - N \sqrt{\beta} \tau ^{k0}   -  \frac{\phi}{N}(  N^k + \beta ^{ik} e ^b _i e _{b0} ) ,\\
&& \phi ^{ij}  = - \frac{1}{2} N \sqrt{\beta} \tau ^{ij}
+\beta ^{ik} \widetilde{\beta} ^{jm} \frac{\phi}{N} e^a _k e^b _m e _{0a} e _{0b}
 .
\end{eqnarray}
Here $N^k=\beta^{ki}N_i$, i.e. we assume that the three-dimensional indices are raised and lowered by the independent metric $\beta^{ik}$ rather than the induced $\widetilde{\beta}_{ik}$.
As a result, embedding matter is described by a set of variables $\phi,\phi ^i,\phi ^{ij}$ and the corresponding contribution to the $S^\text{add}$ action has the form:
\begin{equation}\label{sp4}
S^\text{add}
 =   \int d^4 x
 \left(
(\widetilde{\beta  } _{ij} -\beta _{ij})\phi ^{ij}
 +   (e^a _i e_{a 0} - N_i)\phi ^{i}
+ \left(  N + \frac{1}{N}  e_{a 0} \three{\Pi} _\perp ^{ab} e _{b0}  \right)\phi
 \right) .
\end{equation}
The corresponding GR gravitational degrees of freedom are described by the ADM variables $\beta_{ik}$, $N_k$, $N$.

\section{Transition to canonical variables and constraints}\label{can_vars}
We will construct canonical formalism for the theory with action \eqref{sp3}, in which $S^\text{add}$ is written in the form \eqref{sp4}, and the matter contribution $S_m$ is omitted for simplicity.

First, it is necessary to find expressions for the generalized momenta for each of the independent variables, both for $\beta_{ik}$, $N_k$, $N$ and for $\phi$, $\phi ^i$, $\phi ^{ik}$, $y^ a$.
Since $S^\text{add}$ does not contain time derivatives (hereinafter denoted by a dot) $\dot{N}$, $\dot{N} _k$ and $\dot{\beta} _{ik} $, the generalized momenta $\pi ^{ik},\pi_N ^k,\pi _N$ corresponding to the ADM variables $\beta_{ik},N_k,N$ have the usual form
\begin{equation}
\pi ^{ik} = \frac{\delta S}{\delta \dot{ \beta} _{ik}} = - 2 L ^{ik,lm} K_{lm},\qquad
\pi _N ^k=0,\qquad
\pi _N    =0,
\end{equation}
which gives rise to the standard primary constraints
\begin{equation}\label{ordinary_con}
    \begin{array}{ll}
        \Phi = \pi _N \approx 0, & \qquad \Phi ^k = \pi _N ^k \approx 0.
    \end{array}
\end{equation}
Here and below, the symbol $\approx$ means the fulfilment of the constraint equations in the weak sense, that is, the constraints cannot be assumed to be equal to zero before the calculation of the Poisson brackets.

Note that the velocities $\dot{\phi}$, $\dot{\phi} ^i$, and $\dot{\phi} ^{ik}$ do not appear in the action, which means that the corresponding momenta $\pi^{ \phi}$, $\pi^{\phi}_i$ and $\pi^{\phi}_{ik}$ vanish.
This means the appearance of additional primary constraints:
\begin{equation}\label{DM_constraints}
    \begin{array}{lll}
         \Psi  = \pi^\phi \approx 0; & \qquad
         \Psi _i  = \pi^{\phi}_i \approx 0; & \qquad
         \Psi _{ik}  = \pi^{\phi}_{ik} \approx 0 .
    \end{array}
\end{equation}
For momentum $p _a$ conjugate to the remaining variable $y^a$, we find the expression (note that $\dot{y} ^a=e_0^a$)
\begin{equation}\label{embedding_mom}
    p _a = \frac{\delta S}{\delta \dot{y} ^a} = \phi ^k e _{ak}  + \frac{2}{N}\phi\,  {\three{\Pi} _\perp}{}_{ab}\, \dot{y}^b.
\end{equation}
From this expression one can see that if we assume that $\phi\ne0$, then the transverse part of the velocity vector $\dot{y} ^a$ can be expressed in terms of momenta and coordinates.
For the longitudinal part, this cannot be done, since it does not appear in the equation (\ref{embedding_mom}).
The branch of the solution corresponding to $\phi=0$ at all points $x$ must be studied separately, and we will leave this analysis outside the scope of this paper.
We only note that we can assume that this path corresponds to the possibility to additionally impose Einstein constraints \eqref{sp2} when constructing the canonical formalism, which was mentioned in the Introduction.
Here we also assume that $\phi\ne 0$ is almost everywhere.
The impossibility of expressing the longitudinal part of the velocity from \eqref{embedding_mom} corresponds to the appearance of another primary constraint:
\begin{equation}\label{embedding_constraint}
    \Omega_{j} = p _a e^a _j - \phi  ^k \widetilde{\beta} _{kj}\approx 0.
\end{equation}

We introduce notation that will be useful in what follows.
We will denote
\begin{equation}\label{n_ob1}
p_\perp ^a \equiv \three{\Pi} _\perp  ^{ab} p _b=\frac{2}{N} \phi\,  {\three{\Pi}_\perp}{}^a_b\, \dot{y}^b,
\end{equation}
where \eqref{embedding_mom} is used.
Since both $\dot{y}^b$ and its transverse projection ${\three{\Pi}_\perp}{}^a_b\,\dot{y}^b$ must be timelike vectors, then $ p _a \three{\Pi} _\perp ^{ab} p _b<0$ (recall that the signature $-+++$ is used), so it is convenient to denote
\begin{equation}\label{o1}
p _\perp = \sqrt{- p _a \three{\Pi} _\perp  ^{ab} p _b}.
\end{equation}
One can also note that the zero components of both the vector $\dot{y}^a$ and the vector ${\three{\Pi}_\perp}{}^a_b\,\dot{y}^b$ must be positive, so the quantities $p_\perp ^0$ and $\phi$ must have the same sign, we denote it as
\disn{sp200}{
\dz\equiv\textrm{sign}\,\phi=\textrm{sign}\,p_\perp ^0=\pm 1,
\nom}
where the function $\textrm{sign}\,$ returns the sign of a number.
Note that since $p_\perp ^a$ is a timelike vector, the sign of its component $p_\perp ^0$ can change only at the points where all 7 independent components of the vector $p_\perp ^a$ vanish.
This allows us to assume in a general situation that $p_\perp^0\ne0$, and hence also $\phi\ne0$ for all points $x$.
Thus, the value $\dz=\pm1$ is a global constant, which singles out one of the two possible branches of the solution in the resulting system of Hamilton equations and constraints.
Let us also introduce the normalized transverse velocity vector
\begin{equation}\label{n_ob11}
n^a=\frac{{\three{\Pi}_\perp}{}^a_b\, \dot{y}^b}{\sqrt{- \dot{y}^c {\three{\Pi}_\perp}_{cd} \dot{y}^d}}=\dz\frac{p_\perp ^a}{p_\perp},
\end{equation}
where we use \eqref{embedding_mom} again.

After obtaining expressions for all generalized momenta, one can write down the total Hamiltonian of the theory by doing the Legendre transformation in a standard way and adding all primary constraints with Lagrange multipliers to the result.
The expression for the corresponding Hamiltonian density is obtained by adding primary constraints with Lagrange multipliers to the quantity
\begin{equation}
    \pi^{ik} \dot{\beta} _{ik} + p _a \dot{y} ^a - \mathcal{L},
\end{equation}
where $\mathcal{L}$ is the Lagrangian density.
Note that primary constraints can be used in the expression for the Hamiltonian density.
The resulting Hamiltonian density can be conveniently broken down into two terms
\begin{equation}\label{sp101}
\mathcal{H} = \mathcal{H} ^\text{ADM} + \mathcal{H} ^\text{add},
\end{equation}
where
\begin{equation}
    \mathcal{H} ^\text{ADM} = \frac{1}{2} N \pi ^{ik} \overline{L} _{ik,lm} \pi ^{lm} + \pi ^{ik} \Big( \three{D} _i N_k + \three{D} _k N_i  \Big)  - \frac{1}{2 \varkappa} N \sqrt{\beta} \three{R}+ \lambda \Phi + \lambda _k \Phi ^k
\end{equation}
is a well-known expression from the ADM formalism, and
\begin{equation}\label{h1}
    \mathcal{H} ^\text{add} = \frac{N}{4 \phi} p _a \three{\Pi} _\perp  ^{ab} p _b + \phi ^{ij} (\beta _{ij} - \widetilde{\beta} _{ij} ) + p_a e^a _i \widetilde{\beta} ^{ik} N_k - \phi N
    + \chi \Psi
    + \chi^i \Psi_i
    + \chi^{ij} \Psi_{ij}
    + \xi^i \Omega_i
\end{equation}
corresponds to the embedding matter.
Here $\lambda,\lambda _k,\chi,\chi^{i},\chi^{ij},\xi^i$ are Lagrange multipliers.
Note that in addition to the assumption $N \neq 0$ that is always made, we have also used the above assumption $\phi \neq 0$.

Further, according to Dirac's recipe \cite{dirac_QM} for constructing the canonical description of the theory, it is necessary to require primary constraints to be preserved with time by calculating their Poisson brackets with the total Hamiltonian $H$ and requiring them to vanish.
If the resulting equations cannot be satisfied by the choice of Lagrange multipliers, then this means the appearance of secondary constraints in the theory.

It is easy to show that the conditions for preserving the constraints $\Psi _i$ and $\Omega _i$ are satisfied by fixing the Lagrange multipliers $\xi ^i$ and $\chi ^i$.
The results of calculating the remaining Poisson brackets are:
\begin{eqnarray}
         && \{ H, \Phi  \} = \mathcal{H} _0 ^\text{ADM} - \frac{1}{4\phi} p _\perp ^2 - \phi ; \label{sc1} \\
         && \{ H, \Phi ^k  \} = \beta^{ik} \mathcal{H} _i ^\text{ADM} + \widetilde{\beta} ^{ik} e^a _i p _a; \label{sc2} \\
         && \{ H, \Psi \} = \frac{N}{4 \phi ^2} p _\perp ^2  -N; \label{sc3} \\
         && \{ H, \Psi _{ij} \} = \beta _{ij} -  \widetilde{\beta} _{ij} . \label{sc4}
\end{eqnarray}
Here we use the standard notation for secondary constraints in the ADM formalism for General Relativity:
\begin{eqnarray}
   &&  \mathcal{H} _0 ^\text{ADM} = \frac{1}{2} \pi ^{ik} \overline{L} _{ik,lm} \pi ^{lm} - \frac{1}{2 \varkappa} \sqrt{\beta} \three{R}  ; \label{H0adm}\\
   &&  \mathcal{H} _i ^\text{ADM} = - 2 \beta _{ik} \sqrt{\beta} \three{D} _j \frac{\pi ^{jk}}{\sqrt{\beta}} \label{Hkadm}.
\end{eqnarray}

It is more convenient to use as constraints not the set (\ref{sc1})-(\ref{sc4}) itself, but another, equivalent set.
To do this, we rewrite the expression (\ref{sc3}) as
\begin{equation}
    - \frac{N}{\phi ^2} (\phi - \frac{p _\perp}{2} )(\phi + \frac{p _\perp}{2} ).
\end{equation}
This expression must be equal to zero.
There are two possibilities for this: $\phi = p _\perp/2$ and $\phi = - p _\perp/2$.
To consider them in a unified way, we use the notation \eqref{sp200} introduced above and declare the expression as a constraint
\begin{equation}\label{newcon}
\Xi = \phi - \frac{\zeta}{2} p _\perp \approx 0.
\end{equation}
Thus, we replace the set of constraints (\ref{sc1})-(\ref{sc4}) with the set (\ref{newcon})-(\ref{sc9}):
\begin{eqnarray}
    && \mathcal{H} _0 = \mathcal{H} _0 ^\text{ADM} - \zeta p _\perp \approx 0, \label{sc7} \\
    && \mathcal{H} _k =  \mathcal{H} _k ^\text{ADM} + e^a _k p _a \approx 0,\label{sc8} \\
    && \Sigma _{ij} =  \beta _{ij} -  \widetilde{\beta} _{ij} \approx 0. \label{sc9}
\end{eqnarray}
It is convenient to write the Hamiltonian density \eqref{sp101} as a linear combination of the constraints already introduced:
\begin{equation}\label{Hcon}
\mathcal{H} = N \mathcal{H} _0 + N^k \mathcal{H} _k  + \Sigma _{ij} \big( \phi ^{ij} + \beta ^{ik} \widetilde{\beta} ^{jm} e^a _k p_a N_m \big)
+ \lambda \Phi + \lambda _k \Phi ^k
     + \chi \Psi
    + \chi^i \Psi_i +  \chi^{ij} \Psi_{ij}
    + \xi^i \Omega_i
.
\end{equation}
With this way of writing down the Hamiltonian density, the emerging contribution $-N\Xi ^2/\phi$ was omitted, since the term of the Hamiltonian containing the square of the constraint $\Xi$ does not play a role, since when calculating the secondary constraints and writing out the Hamilton equations, it will always make zero contributions.

Next, one has to check the preservation of secondary constraints.
It is easy to see that the constraint $\Xi$ is preserved by fixing the Lagrange multiplier $\chi$.
Calculating the Poisson brackets of the constraint $\mathcal{H}_i$, one can check that the given constraint is a generator of 3-dimensional diffeomorphisms, which means that its Poisson bracket with the Hamiltonian will be proportional to the constraints and $\mathcal{H} _i$ is preserved automatically.
By a direct and rather cumbersome calculation, one can obtain that in order to preserve the constraints $\mathcal{H} _0$ and $\Sigma _{ij}$, it is necessary to introduce a new constraint
\begin{equation}\label{post_sec_con}
    \Lambda _{ik} =  \overline{L} _{ik,lm} \pi ^{lm} - 2 n _a \three{b}^a _{ik}\approx 0,
\end{equation}
where $\three{b}^a _{ik}$ is the second fundamental form of $x^0=const$ surfaces.
The condition of preserving this constraint generates the new constraint $\Upsilon _{ij}$:
\begin{equation}\label{giant_constraint}
    \Upsilon _{ij} = \{ \Lambda _{ij} , H \}  = \Upsilon _{ij} ^{(1)} - \phi ^{km} \Upsilon _{ij,km} ^{(2)},
\end{equation}
where it is important to note further that the quantities $\Upsilon _{ij} ^{(1)}$ and $\Upsilon _{ij,km} ^{(2)}$ do not depend on $\phi ^{km}$.
The explicit form of these quantities is given in the Appendix.

This completes the list of constraints, since the constraint $\Upsilon _{ij}$ is preserved by choosing the Lagrange multiplier $\chi ^{ij}$.
Summing up this section, we list all the constraints obtained.
Primary constraints are $\Phi$, $\Phi ^k$ (\ref{ordinary_con}); $\Psi$, $\Psi _i$, $\Psi _{ik}$ (\ref{DM_constraints}); $\Omega _j$ (\ref{embedding_constraint}).
Secondary constraints are $\Xi$ (\ref{newcon}); $\mathcal{H} _0$ (\ref{sc7}); $\mathcal{H} _k$ (\ref{sc8}); $\Sigma _{ij}$ (\ref{sc9}); $\Lambda _{ij}$ (\ref{post_sec_con}); $\Upsilon _{ij}$ (\ref{giant_constraint}).
The Lagrange multipliers $\xi ^i$, $\chi$, $\chi ^i$, $\chi ^{ij}$ are fixed.
The Lagrange multipliers $\lambda$ and $\lambda _i$ remain arbitrary.

\section{Constraints classification and their partial solution}\label{classify}
The next step in the Dirac scheme is the classification of all constraints into the first and second class \cite{dirac_QM}.
First class constraints are constraints whose Poisson brackets with all other constraints are just constraints linear combinations.
If a constraint has a non-zero Poisson bracket with any constraint, then it is called a second class constraint.
An important feature of first class constraints is that they can be correctly taken into account when quantizing the theory by imposing them on the state vectors.
Second class constraints must be either solved before quantization, by excluding some pairs of conjugate canonical variables, or Dirac brackets must be introduced.

As already mentioned, the constraint $\mathcal{H}_i$ is the three-dimensional diffeomorphisms generator, which means it is a first class constraint.
Since the theory under consideration is invariant under four-dimensional diffeomorphisms, one should expect that, in addition to $\mathcal{H}_i$, the list of constraints should also contain a first class constraint, corresponding to one more general covariant transformation.
In GR, this is the $\mathcal{H} _0 ^\text{ADM}$ constraint.
In the considered theory, it corresponds to the $\mathcal{H} _0$ constraint (\ref{sc7}). However, not all of its Poisson brackets with other constraints vanish on the constraints surface.
This means that the first class constraint must be some linear combination of $\mathcal{H} _0$ with other constraints.
It is not very useful to look for an explicit form of such a combination since it is much more convenient to simplify the canonical system by solving at least a part of the existing second class constraints.
The same can be said about primary constraints $\Phi$ and $\Phi_k$ \eqref{ordinary_con} (in General Relativity they are first class constraints): one can show that their complicated linear combinations with the constraint $\Psi _{ik}$ \eqref{DM_constraints} also turn out to be first class constraints, but there is no point in obtaining the explicit form of these combinations.

An interesting question is the calculation of the number of degrees of freedom for the resulting Hamiltonian system.
This number can be defined as the number of conjugate canonical variables pairs that remain after eliminating some of the variables by solving \emph{all} of the existing constraints.
Although it is rarely possible to solve all the constraints explicitly, this number can be calculated by subtracting the number of arising equations from the number of initial coordinates and momenta.
It will be more convenient to do this when all first class constraints are already explicitly distinguished from the set of constraints.

To solve constraints means to express any pair of canonical coordinate-momentum variables in terms of other variables and thus eliminate some pair of constraints.
This procedure is directly applicable to second class constraints.
However, if it is required to solve some constraint $T$ belonging to the first class, one can introduce an \emph{additional condition} $\widetilde{T}$ such that $\{ T, \widetilde{T} \} \neq 0$; then some pair of canonical variables can be expressed from the conditions that $T \approx 0 $ and $\widetilde{T} \approx 0$.
Expressed coordinates and momenta must be substituted into the first order action, which in this case has the form
\begin{equation}\label{first_order_action}
    S^{(1)} = \int dt \int d^3 x \Big(  \pi ^{ik} \dot{\beta} _{ik} +  p _a \dot{y} ^a +  \pi _N  \dot{N} +  \pi _N ^k \dot{N} _k +  \pi ^\phi \dot{\phi} +  \pi ^\phi _i \dot{\phi} ^i +  \pi ^\phi _{ij} \dot{\phi} ^{ij} - \mathcal{H} \Big).
\end{equation}
Note that if the momentum in the resolvable pair is equal to zero, then the step with the first order action can be skipped and the solution can be substituted into the Hamiltonian directly.
This happens, for example, with a pair of constraints $\Psi $ and $\Xi $, from which we can express variables
\begin{eqnarray}
&& \phi  = \frac{\dz}{2} p _\perp \label{y1}, \qquad \pi ^\phi = 0,
\end{eqnarray}
where $\dz=\textrm{sign}\,p_\perp ^0$ according to \eqref{sp200}.
Similarly, we can directly substitute into the Hamiltonian the solutions of the $\Psi _i$ and $\Omega _i$ constraint, from which we can express the variables
\begin{eqnarray}
&& \phi ^i= \widetilde{\beta} ^{ik} e^a_k p _a,\qquad
\pi ^\phi_i = 0  ;
\end{eqnarray}
and finally pairs of constraints $\Psi _{ij}$ and $\Upsilon _{ij}$, from which we can express variables
\begin{eqnarray}
&& \phi ^{ik} =  - \Upsilon  _{jl } ^{(1)} \Big( \Upsilon ^{(2) \; -1} \Big) ^{jl,ik},\qquad
 \pi ^\phi _{ik} = 0. \label{y6}
\end{eqnarray}

After solving the listed constraints, the remaining constraints are: $\Phi ^k$, $\Phi$, $\mathcal{H}_k$, $\mathcal{H}_0$, $\Sigma _{ij}$, $\Lambda _{ij}$.
One can verify that now the first class constraints are $\Phi ^k$, $\Phi$, $\mathcal{H}_k$, and the combination
\begin{equation}\label{kj}
   \mathcal{H} _0 ' = \mathcal{H} _0 + \frac{1}{N} \big( \phi ^{ij} + \beta ^{ir} \widetilde{\beta} ^{jq} e^a_r p _a N_q \big) \Sigma _{ij},
\end{equation}
and the Hamiltonian density (\ref{Hcon}) takes the form
\begin{equation}\label{conh2}
\mathcal{H} = N \mathcal{H'} _0 + {N} ^k \mathcal{H} _k+ \lambda \Phi + \lambda _k \Phi ^k.
\end{equation}
This form of the Hamiltonian density, which reduces to a linear combination of eight first class constraints, four of which have the form \eqref{ordinary_con}, exactly corresponds to the situation that takes place in GR.
The difference is the additional presence of the remaining second class constraints $\Sigma _{ij}$ and $\Lambda _{ij}$.

Now it is easy to count the number of degrees of freedom in the system under consideration.
Eliminating $\phi,\phi^{i},\phi^{ik}$ leaves 20 independent variables $\beta_{ik},N_k,N,y^a$.
Together with their conjugate momenta, we have 40 canonical field variables as a result.
At the same time, 8 first class and 12 second class constraints remained (recall that $\Sigma _{ij}$ and $\Lambda _{ij}$ are symmetric $3\times3$ matrices so each have 6 independent components).
For each first class constraint, it is necessary to introduce \emph{additional condition}, so that the total number of equations arising from the constraints is $8*2+12=28$.
Subtracting it from the number of canonical variables, we get $40-28=12$ remaining canonical variables after taking into account the constraints, i.e. 6 pairs.
As a result, we conclude that there are 6 degrees of freedom in the considered system.

Further, it is convenient,
as in the usual ADM formalism, to exclude the pairs of variables $N$, $\pi_N$ and $N_k$, $\pi_N ^k$.
Although the $\Phi$ and $\Phi ^k$ constraints are first class constraints, they can be solved by adding additional conditions (actually, these are gauge conditions) on $N$ and $N_k$.
We write these conditions as $\widetilde{\Phi} = N- \widetilde{N} \approx 0$ and $\widetilde{\Phi} _k = N_k - \widetilde{N} _k \approx 0$, where $\widetilde{N}$, $\widetilde{N} _k$ are arbitrary functions.
The choice of some specific functions as $\widetilde{N}$, $\widetilde{N} _k$ would mean the use of some specific gauge.
As a result, there remains only the $\mathcal{H'} _0$ (\ref{kj}), $\mathcal{H}_k$ (\ref{sc8}), $\Sigma _{ij}$ (\ref{sc9}) and $\Lambda _{ij}$ (\ref{post_sec_con}) constraints, and the Hamiltonian (\ref{conh2}) density is reduced to the expression
\begin{equation}\label{conh2a}
\mathcal{H} = \widetilde{N} \mathcal{H'} _0 + \widetilde{N} ^k \mathcal{H} _k.
\end{equation}
In what follows, instead of $\widetilde{N}$, $\widetilde{N}^k$ we will write $N$, $N^k$, i.e. by $N$ and $N^k$ we will now mean arbitrary functions, which are usually the Lagrange multipliers.

After eliminating the variables $N$, $\pi_N$ and $N_k$, $\pi_N ^k$, four first class constraints $\mathcal{H} _k$ and $\mathcal{H} _0 '$ remain, as well as $ \Sigma _{ij}$ and $\Lambda _{ij}$.
Since $\{ \Sigma _{ij},\Lambda _{km} \} \not \approx 0$, the latter are still second class constraints.
If they are solved, then only the first class constraints will remain in the canonical system.
There are two possibilities for solving the constraints $\Sigma _{ij}$ and $\Lambda _{ik}$: either use them to eliminate the embedding function $y^a$ and its conjugate momentum $p_a$, leaving the metric $\beta _{ij}$ and its conjugate momentum $\pi^{ij}$ as independent variables, or vice versa, eliminate $\beta _{ij}$ and $\pi^{ij}$.
The first option is an interesting problem from the point of view of studying the embedding matter properties, but it is very difficult to solve.
This is due to the fact that the constraint $\Sigma _{ij}$ \eqref{sc9}, understood as an equation on $y^a$, is a non-linear partial differential equation:
\begin{equation}
(\partial _i y^a)(\partial _k y_a) = \beta _{ik}.
\end{equation}
As an alternative to this method of solving second class constraints, one can try to use the formalism of Dirac brackets \cite{dirac_QM}. However, such approaches are beyond the scope of this paper.
Instead, we explore the possibility of solving constraints in a second, simpler way, eliminating the variables $\beta _{ij}$ and $\pi^{ij}$.

\section{Solving the remaining second class constraints}\label{compare}
Let's solve pair of constraints $\Sigma _{ij}$ (\ref{sc9}) and $\Lambda _{ij}$ (\ref{post_sec_con}) with respect to variables $\beta _{ij}$ and $\pi^{ ij}$:
\begin{equation}\label{resh}
    \begin{array}{ll}
\beta _{ij} = \widetilde{\beta} _{ij},  & \qquad \pi ^{ij} =  2 n_a \three{b}^a _{lk} L^{lk,ij} .
    \end{array}
\end{equation}
We substitute the resulting solution into the first order action (\ref{first_order_action}).
Since some constraints have already been solved, it takes the form
\begin{equation}\label{first_order_action2}
S^{(1)} = \int dt \int d^3 x \Big(  \pi ^{ij} \dot{\beta} _{ij} +  p _a \dot{y} ^a  - N \mathcal{H}_0 - N^k \mathcal{H} _k   \Big),
\end{equation}
where the form (\ref{conh2a}) of the Hamiltonian density is used and it is taken into account that one can replace $\mathcal{H'}_0$ by $\mathcal{H}_0$ when the constraint $\Sigma _{ij}$ is implied.

Consider separately the contribution of the term $\pi ^{ij} \dot{\beta} _{ij}$.
After substituting $\beta _{ij}$ in accordance with (\ref{resh}), we replace the ordinary derivative $\partial_i$ with the covariant $\three{D} _i$ and integrate by parts:
\begin{equation}\label{wx}
 \int d^3 x\,   \pi ^{ij} \dot{\beta} _{ij}
 = -2 \int d^3 x  \ls \sqrt{\beta}\,\dot y^a  e_{aj}  \three{D} _i \frac{\pi ^{ij} }{\sqrt{\beta}}+
 \pi ^{ij}  \dot y^a \three{D} _i e_{aj}\rs.
\end{equation}
Note that in this expression the first term can be rewritten in terms of $\mathcal{H} _i ^\text{ADM}$ using (\ref{Hkadm}), and the second term contains the second fundamental form of the surface $\three{b} _ {aij} = \three{D}_i e _{aj}$.
Expressing $\pi ^{ij}$ according to (\ref{resh}) we get
\begin{equation}
 \int d^3 x\,\pi ^{ij} \dot{\beta} _{ij}  =
 \int d^3 x\,\Big( \mathcal{H} ^\text{ADM} _i \widetilde{\beta} ^{ij}  e_{aj} - 4 n_c \three{b} ^c _{lk} L^{lk,ij} \three{b} _{aij} \Big) \dot{y} ^a.
\end{equation}
We emphasize that the quantities $\beta _{ij}$ and $\pi^{ij}$ appearing in this expression (the latter appearing via $\mathcal{H} ^\text{ADM} _k$) are considered to be expressed via $y^ a$ and $p_a$ according to (\ref{resh}).

We substitute the resulting expression into \eqref{first_order_action2}, introducing the notation
\begin{equation}\label{sp105}
B^{ab} = -4 \three{b} ^a _{ik}  L^{ik,lm}\three{b} ^b _{lm}=
-\frac{\sqrt{\beta}}{2\varkappa}\three{b} ^a _{ik} \three{b} ^b _{lm}  \Big( \beta ^{il} \beta ^{km} + \beta ^{im} \beta ^{kl} - 2 \beta ^{ik} \beta ^{lm} \Big)
\end{equation}
and using the constraint (\ref{sc8}) to replace $\mathcal{H}_i^\text{ADM}$ with $-e^a_i p_a$.
Such a change is equivalent to some change in the Lagrange multiplier $N^k$, which is insignificant because of its arbitrariness.
As a result, we get
\begin{equation}\label{sp102}
S^{(1)} = \int d^4 x \ls\ls B_{ab} n^b + p_{\perp a} \rs \dot{y} ^a - N \mathcal{H} _0 - N^k \mathcal{H} _k\rs.
\end{equation}
Further, we note that, according to \eqref{n_ob11}, we can write $p_{\perp a} = \dz p_\perp n_a$, which, using the relation (\ref{sc7}) (at which the Lagrange multiplier $N$ now insignificantly changes) allows us to replace $p_{\perp a}$ in \eqref{sp102} with $- \mathcal{H} _0 ^\text{ADM} n_a$.
In turn, the quantity $\mathcal{H} _0 ^\text{ADM}$, as can be shown (see \cite{statja18}), in terms of the variables $y^a$ and $p_a$ is written as
\begin{equation}
\mathcal{H} _0 ^\text{ADM}=-\frac{1}{2} (n_c B ^{cb} n_b + B^c _c).
\end{equation}
This allows to rewrite the first order action \eqref{sp102} in the final form
\begin{equation}\label{sp103}
S^{(1)} = \int d^4 x \ls\Big( B_{ab} n^b  + \frac{1}{2} (n_c B ^{cb} n_b + B^c _c) n_a \Big) \dot{y}^a  - N \mathcal{H} _0 - N^k \mathcal{H} _k\rs.
\end{equation}

Now we need to construct the Hamiltonian formulation for the theory given by such an action, following the same Dirac scheme that was used for the original theory in the previous sections.
In this case, it is necessary to consider all the fields included in the action \eqref{sp103} as generalized coordinates, i.e. $y^a$, $p_a$, $N$, and $N_k$ (it is the variation with respect to the last two that leads to the appearance of the constraints $\mathcal{H} _0$ and $\mathcal{H} _k$ among the equations of motion), introducing the associated generalized momentum for each of them.
Next, we need to find and classify constraints, and then solve those of them for which this is possible.

First, let us write out the expressions for the new generalized momenta:
\disn{sp106}{
\pi _a = \frac{\delta S ^{(1)}}{\delta \dot{y} ^a} = B_{ab} n^b  + \frac{1}{2} (n_c B ^{cb} n_b + B^c _c) n_a,
\nom}\vskip -2em
\disn{sp107}{
\pi _p ^a = \frac{\delta S ^{(1)}}{\delta \dot{p} _a} = 0,\qquad
\pi _N = \frac{\delta S ^{(1)}}{\delta \dot{N}} = 0,\qquad
\pi _N ^k = \frac{\delta S ^{(1)}}{\delta \dot{N} _k} = 0.
\nom}
We obtain primary constraints from the expression for $\pi_a$, taking into account the definition \eqref{sp105} (note that the second fundamental form $\three{b}^a_{ik}$ is always transverse in its index $a$) and the fact that $n^a$ is normalized (see \eqref{n_ob11}):
\disn{sp108}{
\pi _a e^a _i \approx 0,\qquad
n (y^a, \pi _a) ^2 + 1 \approx 0.
\nom}
Here $n (y^a, \pi _a)$ is an implicitly defined function that is a solution to \eqref{sp106} considered as a cubic equation on $n^a$; see \cite{statja44} for details.
From \eqref{sp106}, as well as from \eqref{sp107}, other, independently of \eqref{sp108}, primary constraints follow, which can be written as
\disn{sp109}{
n^a (y^a, \pi _a) - \dz\frac{p_\perp ^a}{p_\perp} \approx 0,
\nom}\vskip -2em
\disn{sp109a}{
\pi _p ^a \approx 0,\qquad
\pi _N \approx 0,\qquad
\pi _N ^k \approx 0.
\nom}
Next, we write an expression for the new Hamiltonian of the theory, which reduces to a linear combination of $\mathcal{H}_0$ \eqref{sc7} and $\mathcal{H}_k$ \eqref{sc8} with coefficients $N$ and $ N^k$ respectively, and first class constraints \eqref{sp108}-\eqref{sp109a} with new Lagrange multipliers.

From the requirement to preserve the $\pi _N \approx 0$ and $\pi _N ^k \approx 0$ constraints, we obtain secondary constraints
\disn{sp110}{
\mathcal{H} _0 \approx 0, \qquad \mathcal{H} _k \approx 0,
\nom}
and from the requirement to preserve the $\pi_p ^a\approx 0$ constraint, we obtain secondary constraints
\disn{sp111}{
N\approx 0, \qquad N_k\approx 0.
\nom}
The appearance of the last two constraints is since the longitudinal part of the variable $p^a$ enters the Hamiltonian only through $\mathcal{H}_k$, and the modulus of its transverse part $p_\perp$ enters only through $\mathcal{H}_0 $.
Further analysis shows that no other secondary constraints arise.

As a further step, the easiest way is to solve some constraints by eliminating the pairs of variables $N$, $\pi _N$; $N_k$, $\pi _N ^k$; $p_a$, $\pi ^a _p$.
Since, according to \eqref{sp109a}, all generalized momenta in these pairs are equal to zero, the expressed variables can be substituted directly into the Hamiltonian, avoiding the need to consider the first order action again.
The generalized coordinates of these pairs of variables are expressed from \eqref{sp111} and the relation
\disn{sp112}{
p^a=-e^a_i\be^{ik}\mathcal{H} _k ^\text{ADM}+n^a(y^a, \pi _a)\mathcal{H} _0 ^\text{ADM},
\nom}
which follows from \eqref{sp109} and \eqref{sp110} given \eqref{sc7},\eqref{sc8}.

As a result, only the constraints \eqref{sp108} remain unresolved, and the final expression for the Hamiltonian density reduces to their linear combination with Lagrange multipliers:
\disn{sp113}{
\mathcal{H} = \la^i e^a _i\pi_a+ \la^4\ls n (y^a, \pi _a) ^2 + 1 \rs.
\nom}
This expression for the Hamiltonian density exactly coincides with the one found in \cite{statja44} when constructing a canonical formalism for a complete embedding theory described only by the embedding function $y^a$.
Thus, if within the framework of the embedding theory in the from of GR with additional matter the Einstein variables $\be_{ij}$, $\pi^{ij}$ are eliminated when solving the last pair of second class constraints (\ref{sc9}),(\ref{post_sec_con}), then we obtain the canonical formulation of the complete embedding theory with an implicitly defined constraint known from \cite{statja44}.
This fact confirms the consistency of the approach used and is also a verification of the results obtained in the previous sections.

After the additional solution of the constraints, we can return to the question already discussed in the \ref{classify} section about counting the number of degrees of freedom in the system under consideration.
The independent variables are now only 10 components of $y^a$, which together with conjugate momenta give 20 canonical field variables.
The remaining constraints are 4 constraints \eqref{sp108}. They are first class constraints, this is shown in \cite{statja44}, where the exact form of the algebra formed by these constraints is obtained.
Taking into account the introduction of \emph{additional conditions} for first class constraints, the number of equations arising from the constraints is $4*2=8$.
Subtracting it from the number of canonical variables, we get $20-8=12$ remaining canonical variables after taking into account the constraints, i.e., as before, 6 pairs, which means 6 degrees of freedom.
In addition to the two degrees of freedom corresponding to General Relativity, we see 4 more degrees of freedom corresponding to embedding matter.
The $\tau^{0\mu}$ components of the embedding matter energy momentum tensor, which are related to its density and velocity, can be considered as such independent degrees of freedom.
The remaining components $\tau^{ik}$ are expressed in terms of $\phi^{ik}$ from \eqref{spn1} and are found when solving the relation \eqref{y6}.
This corresponds to some complex embedding matter equation of state.

\section{Conclusions}
For the complete embedding theory formulated in the form of GR with an additional contribution of the so-called \textit{embedding matter} \cite{statja48,statja51}, the canonical (Hamiltonian) description of the theory is constructed.
All the constraints arising from this approach are found: primary (\ref{ordinary_con}), (\ref{DM_constraints}), (\ref{embedding_constraint}) and secondary (\ref{newcon})-(\ref{sc9}), (\ref{post_sec_con}), (\ref{giant_constraint}), as well as the representation of the Hamiltonian $H$ in terms of constraints (\ref{Hcon}).
Most of the arising second class constraints can be solved, as well as simple first class constraints with the introduction of additional conditions.
As a result, we get the canonical formulation of the theory with the  Hamiltonian density \eqref{conh2a}, which reduces to the linear combination of constraints \eqref{sc8},\eqref{kj}, and with second class constraints (\ref{sc9}), (\ref{post_sec_con}).

The obtained canonical description of the complete embedding theory can be used both in attempts to quantize the theory and in the analysis of its classical equations of motion to study the embedding matter properties.
A comparison of the obtained properties with the properties of dark matter known from observations is necessary to understand whether the effects explained by the presence of dark matter can be explained within the framework of replacing General Relativity with the embedding theory as the theory of gravitational interaction.
To advance in each of these two directions, it is necessary to take into account correctly the presence of the remaining second class constraints (\ref{sc9}),(\ref{post_sec_con}) in the theory.
From the point of view of obtaining embedding matter properties, it would be most useful to solve these constraints by excluding canonical variables in such a way that the variables $\be_{ik}$, $\pi^{ik}$ associated with the metric, as well as variables describing embedding matter remain.
However, finding a way to solve constraints in such a way remains an open problem and requires additional study, since the constraint (\ref{sc9}) is a non-linear differential equation for the variable $y^a$.
As an alternative, one can try to avoid the direct solution of second class constraints by using the Dirac bracket formalism.

Another option for taking into account these second class constraints is to solve them by eliminating the variables $\be_{ik}$, $\pi^{ik}$.
Although this is less interesting from the point of view of analyzing the embedding matter properties, it allows the independent verification of the consistency of the embedding theory description in the form of GR with embedding matter.
As we have shown, in this way, the canonical system constructed in this paper passes into the canonical formulation of the complete embedding theory with an implicitly defined constraint, which was obtained in \cite{statja44}.

An interesting question is the connection between the obtained canonical system and the canonical description of the embedding theory equivalent to GR with the additional imposition of Einstein constraints $\mathcal{H}_0 ^\text{ADM}\approx0$, $\mathcal{H}_k^\text{ ADM}\approx0$ \cite{regge,statja18,statja24}.
As can be seen from \eqref{sc7}, \eqref{sc8}, imposition of these constraints leads to both the longitudinal part of the momentum $e^a_k p_a\approx0$ and the transverse part $p_\perp\approx0$ be equal to zero.
The latter, given \eqref{o1} and \eqref{n_ob1}, and the fact that $\dot{y}^a \three{\Pi}_{\perp ab} \dot{y}^b< 0$ leads to the condition $\phi \approx 0$, which contradicts our earlier assumption $\phi \neq 0$, see the comment after the formula (\ref{embedding_mom}).
Thus, the Einstein limit of the considered canonical formalism turns out to be singular and requires a separate study.

{\bf Acknowledgements.}
The work is supported by RFBR Grant No.~20-01-00081.

\section*{Appendix: explicit form of the $\Upsilon _{ij}$ constraint}
The explicit form of the $\Upsilon _{km}$ constraint is as follows:
\begin{equation}\label{giant_constraint2}
    \Upsilon _{km}  = \Upsilon _{km} ^{(1)} - \phi ^{ij} \Upsilon _{km,ij} ^{(2)},
\end{equation}
where
\disn{spn1}{
    \Upsilon _{km} ^{(1)} = \Big( L _{km,ij} + 4 \zeta \three{b} ^a _{km} \frac{\three{\Pi} ^c _{a \perp} + n_a n^c}{p_\perp} \three{b} _{cij} \Big) \beta ^{ir} \widetilde{\beta} ^{jq} e^b_r p _b N_q
     -  2 \zeta  \frac{\three{b} ^a _{km}}{p _\perp} \partial _q (N n _a p _b e^{bq})
  +\ns+ \big( 2 n_c e^q _b \partial _k \partial _m y^b  + 2 \zeta \frac{p _a  \three{b} _{kmc}}{p _\perp} e^{aq}  \big) \partial _q (N n^c)
  + 2 n_a
    \partial _k \partial _m ( N n^a )
+  2 \zeta \frac{n _a \three{b} ^a _{km}}{p_\perp} p_b  e^{bq} \partial _q N
+\ns+
2 \zeta \frac{n _a \three{b} ^a _{km}}{p_\perp} \partial _q (N p_b  e^{bq})
+ \frac{4 \varkappa}{\sqrt{\beta}}  N  \pi ^{lq} \pi ^{ij}   \Big(
        \beta _{mj} \overline{L} _{ki,lq}   + \beta _{ki}  \overline{L} _{mj,lq}
      - \frac{3}{8} \beta _{km}  \overline{L} _{ij,lq}
   + \frac{1}{2}  \beta _{lq}  \overline{L} _{km,ij} - \frac{1}{2} \beta _{qj} \overline{L} _{km,li}
    \Big)
    - \ns - \frac{N  \sqrt{\beta}}{2\varkappa} \overline{L} _{km,rs}   \three{G} ^{rs}
    + 4
       \three{D} _k \three{D} _m N ,
\nom}
and
\begin{equation}
    \Upsilon _{km,ij} ^{(2)} = L _{km,ij} - 4 \zeta \three{b} ^a _{km} \frac{\three{\Pi} ^c _{a \perp} + n_a n^c}{p_\perp} \three{b} _{cij} .
\end{equation}


\end{document}